\documentclass[seceq]{ptptex}
\usepackage{bm}
\usepackage{wrapft}
\usepackage[unicode=true,
 bookmarks=false,
 breaklinks=false,pdfborder={0 0 1},backref=false,colorlinks=false]
 {hyperref}
\hypersetup{
 bookmarksnumbered=false,bookmarksopen=false}

\title{Mass splittings of SU(3) baryons within a chiral soliton model}
\author{Ghil-Seok \textsc{Yang}$^1$\footnote{E-mail:
    ghsyang@knu.ac.kr} and Hyun-Chul
  \textsc{Kim}$^{2,3,4}$\footnote{E-mail: hchkim@inha.ac.kr}} 
\inst{$^1$Center for High Energy Physics \& Department of Physics,
  Kyungpook National University, Daegu 702-701, Republic of Korea\\
$^2$Department of Physics, Inha University, Incheon 402-751,
  Republic of Korea \\
$^3$Department of Physics, University of Connecticut,
  Storrs, CT 06269, U.S.A. \\
$^4$School of Physics, Korea Institute for Advanced Study,
  Seoul 130-722, Republic of Korea}

\date{April 2012}
\abst{
Considering simultaneously isospin and SU(3) flavor symmetry breakings,
we investigate the complete mass splittings of SU(3) baryons within
a chiral soliton model, a {}``model-independent approach'' being
employed. In linear order several new mass relations are derived,
which are mostly generalizations of existing mass formulae. The dynamical
quantities appearing in the expressions for the masses are fixed by
fitting them to the masses of the baryon octet and those of
$\Omega^{-}$ and $\Theta^{+}$ as input rather than by extracting them
from a calculated self-consistent soliton profile. In particular, the
consideration of isospin symmetry breaking allows us to use the
experimental data of the whole octet baryon masses as input. We
predict the masses of the baryon decuplet and antidecuplet without any
further adjustable free parameter. In addition, we also obtain the
pion-nucleon sigma term which turns out to be $\Sigma_{\pi
  N}\,=\,36.4\pm3.9\,\mathrm{MeV}$. We get the ratio of the current
light quark masses $R\;=\;58.1\pm1.3$. The present results indicate
that the recent experimental data for the $\Theta^{+}(1524)$ are
compatible with the experimental data of the octet and decuplet
masses.  }

\begin{document}
\maketitle

\section{Introduction}

The mass splittings of SU(3) baryons are the first observables that
any low-energy effective model for quantum chromodynamics (QCD) should
explain. Since the Skyrme model was suggested as a topological effective
model of QCD in the large $N_{c}$ limit, the picture of the topological
and particularly non-topological chiral soliton has been known to
be very successful in describing the splitting between the baryon
octet and decuplet. One of the most interesting features in this picture
is that the low-lying baryons can be regarded as rotational excitations
of the chiral soliton: The baryon octet ($\bm{8}$) appear as the
lowest representation with spin $1/2$ and positive parity. The baryon
decuplet ($\bm{10}$) with spin $3/2$ arises from the next rotational
excitation of the chiral soliton. When we proceed further with these
excitations, we find that there are baryon antidecuplet ($\overline{\bm{10}}$),
eikosiheptaplet ($\bm{27}$), and so on. In particular, the baryon
antidecuplet is the first excitation consisting of exotic pentaquark
baryons~\cite{Jezabek:1987ns,Diakonov:1997mm,Praszalowicz:2003ik}
which have attracted much attention, since the LEPS collaboration
announced the first measurement of the pentaquark baryon
$\Theta^{+}$~\cite{Nakano:2003qx}. However, a series of the CLAS
experiments has reported null results of the
$\Theta^{+}$~\cite{Battaglieri:2005er,McKinnon:2006zv,Niccolai:2006td,DeVita:2006ha}
and has casted doubt on its existence. On the other hand, the DIANA
collaboration has continued to search for the
$\Theta^{+}$~\cite{Barmin:2006we,Barmin:2009cz} and announced very
recently the formation of a narrow $pK^{0}$ peak with mass of
$1538\pm2$ MeV/$c^{2}$ and width of $\Gamma=0.39\pm0.10$ MeV in the
$K^{+}n\rightarrow K^{0}p$ reaction with higher statistical
significance ($6\,\sigma-8\,\sigma$)~\cite{Barmin:2009cz}. Moreover,
other new positive experiments for the $\Theta^{+}$ have been
reported~\cite{new_SVD,Hotta:2005rh,Miwa:2006if,SVD2:2008}. The LEPS
collaboration also reported the evidence of the $\Theta^{+}$
existence~\cite{Nakano:2008ee}:
$M_{\Theta}=1.524\pm0.002\pm0.003\,\mathrm{GeV}/c^{2}$ with the
statistical significance $5.1\,\sigma$. The peak position is shifted
by $+3$ MeV systematically due to the minimum momentum spectator
approximation. 

In addition to the $\Theta^{+}$ baryon, the GRAAL
experiment~\cite{Kuznetsov:2004gy,Kuznetsov:2006de,Kuznetsov:2006kt}
has found a new nucleon-like resonance around $1.67$ GeV from $\eta$
photoproduction off the deutron in the neutron channel. The decay
width was measured to be around $40$ MeV. However, Fermi-motion being
effects excluded, this decay width may decrease~\cite{Fix:2007st}.
References~\cite{Kuznetsov:2004gy,Kuznetsov:2006de,Kuznetsov:2006kt}
have shown that the resonant structure was not seen in the quasi-free
proton channel. Note that this newly found $N^{*}$ resonance is consistent
with the theoretical predictions~\cite{Diakonov:2003jj,Arndt:2003ga}
of non-strange exotic baryons. Moreover, the narrow width and its
dependence on the initial isospin state are the typical characteristics
for the photo-excitation of the non-strange antidecuplet
pentaquark~\cite{Polyakov:2003dx,Kim:2005gz}. Very recently, a new
analysis of the free proton GRAAL
data~\cite{Kuznetsov:2007dy,Kuznetsov:2008gm,Kuznetsov:2010as,Kuznetsov:2008hj,Kuznetsov:2008ii} 
has revealed a resonance structure with a mass around 1685 MeV and
width $\Gamma\leq15$ MeV. However, we have to mention that the results
of Ref.~\cite{Kuznetsov:2007dy} do not agree with those of
Ref.~\cite{Bartalini:2007fg}. For a detailed discussion of this
discrepancy, we refer to Ref~\cite{Kuznetsov:2008gm}. The CB-ELSA
collaboration~\cite{CBELSA} has also announced an evidence 
for this newly found $N^{*}$ resonance in line with that of GRAAL.
All these experimental facts are consistent with the results for the
transition magnetic moments in the $\chi$QSM~\cite{Polyakov:2003dx,Kim:2005gz}
and phenomenological analysis for the non-strange pentaquark
baryons~\cite{Azimov:2005jj}. Based on these results, theoretical
calculations of the $\gamma N\to\eta N$
reaction~\cite{Choi:2005ki,Choi:2007gy} describe qualitatively well
the GRAAL data. In Refs.~\cite{Arndt:2003ga,non_strange_partner2} the
non-strange partners of the $\Theta^{+}$ were also studied, results of
which are comparable with those in this work. 

In the original predictions of the $\Theta^{+}$ mass from a chiral
soliton model ($\chi$SM)~\cite{Diakonov:1997mm} the theoretical
analyses are partially based on specific model
calculations~\cite{Christov:1993ny,Blotz:1994wi}, while some dynamical
parameters are fixed by the experimental masses of the baryon octet
and decuplet and the empirical value of the $\pi N$ sigma term
$\Sigma_{\pi N}$. In particular, the second moment of
inertia~\cite{Blotz:1992br,Blotz:1992pw} of the chiral soliton known
as $I_{2}$, which is an essential quantity to determine the shift of
the antidecuplet center from the octet center in the chiral limit, is
given only in a wide range: $0.43\,\mathrm{fm}\,<\,
I_{2}\,<0.55\,\mathrm{fm}$, 
depending on specific models such as either the Skyrme
model~\cite{Walliser:1992am,Walliser:1992vx} or the chiral
quark-soliton model
($\chi$QSM)~\cite{Blotz:1992br,Blotz:1992pw}. Thus, some of
model-dependent uncertainties are inherent in previous analyses of the
SU(3) baryon masses. 

In the present work, we aim at determining the masses of the baryon
decuplet, and a part of those of the baryon antidecuplet with all
dynamical parameters fixed to existing data for the masses of the
baryon octet~\cite{PDG}. In order to incorporate these data, however,
it is essential to consider the breakdown of isospin symmetry, since
the experimental data involve already the effects of isospin symmetry
breaking that consist of two different contributions: The electromagnetic
(EM) and hadronic ones. The isospin breaking effects due to the EM
corrections were already investigated in Ref.~\cite{Yang:2010id}
within the framework of the $\chi$SM. In the present work, we will
introduce additionally the hadronic contributions of isospin symmetry
breaking to the mass splittings, so that we can analyze the mass splittings
of the SU(3) baryons consistently.

Having taken into account these effects of isospin symmetry breaking,
we are able to fix unequivocally the relevant model parameters by
employing the experimental data of the baryon octet masses. In
addition, we will use the experimental values of the
$\Theta^{+}(1524)$, though its existence is disputable, 
as well as of the $N^{*}(1685)$ such that we can determine $I_{2}$
unambiguously. We will show that the masses of the baryon decuplet
and parts of the antidecuplet are determined uniquely without any
adjustable parameters.

The present work is sketched as follows: In Section II, we describe
the general formalism of the present approach for the mass splittings
of the SU(3) baryons. We also discuss various relations between SU(3)
baryon masses such as the generalized relations of Gell-Mann-Okubo,
of Coleman-Glashow relation, and of Guadagnini. In Section III, we
show how to determine the dynamical parameters of the $\chi$SM, using
the existing experimental data for the octet with $\Omega^{-}(1672)$
and $\Theta^{+}(1524)$ masses. We present the final results of the
decuplet and antidecuplet masses. In the last Section, we summarize
the present work and draw conclusions.

\section{General Formalism}

The mass splittings of SU(3) baryons within a chiral soliton have
been extensively studied in Refs.~\cite{Diakonov:1997mm,Ellis:2004uz}.
As mentioned in the previous Section, the effects of isospin symmetry
breaking arise from two different sources, i.e. the mass differences
of the up and down quarks and the EM interactions. The effects of
isospin symmetry breaking on the baryon mass splittings have been
studied in Refs.~\cite{Praszalowicz:1992gn,Blotz:1994pc} within
the $\chi$QSM. The EM mass splittings of the SU(3) baryons have been
already investigated in Ref.~\cite{Yang:2010id}. Thus, we briefly
review how to construct the collective Hamiltonians for the masses
of the SU(3) baryons with isospin symmetry breaking taken into account
in the present Section.

\subsection{Collective Hamiltonian and SU(3) baryon states}

We start from the collective Hamiltonian in the SU(3)
$\chi\mathrm{SM}$~\cite{Blotz:1992pw,Blotz:1992br}: 
\begin{eqnarray}
H & = & M_{{\rm cl}}\;+\; H_{\mathrm{rot}}\;+\; H_{\mathrm{sb}},
\label{eq:collH}
\end{eqnarray}
 where $M_{\mathrm{cl}}$ denotes the classical soliton mass. The
$H_{\mathrm{rot}}$ and $H_{\mathrm{sb}}$ respectively stand for
the $1/N_{c}$ rotational and symmetry-breaking corrections including
the effects of isospin and $\mathrm{SU(3)_{f}}$ symmetry
breakings~\cite{Blotz:1994pc}: 
\begin{eqnarray}
H_{\mathrm{rot}} 
& = & 
\frac{1}{2I_{1}}\sum_{i=1}^{3}\hat{J}_{i}^{2}
\;+\;\frac{1}{2I_{2}}\sum_{p=4}^{7}\hat{J}_{p}^{2},
\label{eq:rotH}
\end{eqnarray}
\begin{eqnarray}
H_{\mathrm{sb}} 
& = & 
\left(m_{\mathrm{d}}-m_{\mathrm{u}}\right)\left(\frac{\sqrt{3}}{2}
\,\alpha\, D_{38}^{(8)}(\mathcal{R})\;+\;\beta\,\hat{T_{3}}
\;+\;\frac{1}{2}\,\gamma\sum_{i=1}^{3}D_{3i}^{(8)}(\mathcal{R})\,\hat{J}_{i}\right)
\cr
 &  & +\;\left(m_{\mathrm{s}}-\hat{m}\right)\left(\alpha\,
   D_{88}^{(8)}(\mathcal{R})
\;+\;\beta\,\hat{Y}\;+\;\frac{1}{\sqrt{3}}
\,\gamma\sum_{i=1}^{3}D_{8i}^{(8)}(\mathcal{R})\,\hat{J}_{i}\right)\cr
 &  & +\;\left(m_{u}+m_{d}+m_{s}\right)\sigma,
\label{eq:sbH}
\end{eqnarray}
 where $I_{1,2}$ represent the soliton moments of inertia that depend
on dynamics of specific formulations of the $\chi$SM. The $J_{i}$
denote the generators of the SU(3) group. The $m_{\mathrm{u}}$,
$m_{\mathrm{d}}$, and $m_{\mathrm{s}}$ designate the up, down, and
strange current quark masses, respectively. The $\hat{m}$ is the
average of the up and down quark masses. The
$D_{ab}^{(\mathcal{R})}(\mathcal{R})$ indicate the SU(3) Wigner $D$
functions. The $\hat{Y}$ and $\hat{T_{3}}$ are the operators of the
hypercharge and isospin third component, respectively. The $\alpha$,
$\beta$, and $\gamma$ are given in terms of the $\pi N$ sigma term
$\Sigma_{\pi N}$ and soliton moments of inertia $I_{1,2}$ and
$K_{1,2}$ as follows:  
\begin{equation}
\alpha=-\left(\frac{2}{3}\frac{\Sigma_{\pi N}}{m_{\mathrm{u}}
+m_{\mathrm{d}}}-\frac{K_{2}}{I_{2}}\right),\;\;\;\;
\beta=-\frac{K_{2}}{I_{2}},\;\;\;\;
\gamma=2\left(\frac{K_{1}}{I_{1}}-\frac{K_{2}}{I_{2}}\right).
\label{eq:abg}
\end{equation}
 Since $\alpha$, $\beta$, and $\gamma$ depend on the moments of
inertia, they are also related to details of specific dynamics of
the $\chi\mathrm{SM}$. Note that $\alpha$, $\beta$, and $\gamma$
defined in the present work do not contain the strange quark mass,
while those in Refs.~\cite{Diakonov:1997mm,Ellis:2004uz} include
it. The $\sigma$ is proportional to the $\Sigma_{\pi N}$ as follows:
\begin{equation}
\sigma\;\;=\;\;-(\alpha+\beta)
\;\;=\;\;\frac{2}{3}\frac{\Sigma_{\pi
    N}}{m_{\mathrm{u}}+m_{\mathrm{d}}},
\label{eq:sigma}
\end{equation}
 which can be absorbed by the center of the mass splittings from the
rotational Hamiltonian $H_{\mathrm{rot}}$.

In the $\chi\mathrm{SM}$, there is a very important constraint for
the collective quantization: 
\begin{equation}
J_{8}\;=\;-\frac{N_{c}}{2\sqrt{3}}B
\;=\;-\frac{\sqrt{3}}{2},\;\;\;\; Y'
\;=\;\frac{2}{\sqrt{3}}J_{8}
\;=\;-\frac{N_{c}}{3}\;=\;-1,
\label{eq:constraintJ}
\end{equation}
 where $B$ is the baryon number. It is related to the eighth component
of the soliton angular velocity that is due to the presence of the
discrete valence quark level in the Dirac-sea spectrum in the SU(3)
$\chi\mathrm{SM}$~\cite{Blotz:1992pw,Christov:1995vm}, while it
arises from the Wess-Zumino term in the SU(3) Skyrme model~
\cite{Witten:1983tx,Guadagnini:1983uv,Jain:1984gp}. Its presence has
no effects on the chiral soliton but allows us to take only the
$\mathrm{SU(3)_{f}}$ irreducible representations with zero
triality. Thus, the allowed $\mathrm{SU(3)_{f}}$ multiplets are the
baryon octet ($J=1/2$), decuplet ($J=3/2$), and antidecuplet
($J=1/2$), etc. In the representation $(p,\, q)$ of the SU(3) group,
we can have the following relation:
\begin{equation}
\sum_{i=1}^{8}J_{i}^{2}
=\frac{1}{3}\left[p^{2}
\;+\; q^{2}\;+\; p\;
  q\;+\;3(p+q)\right],
\label{eq:Jsquare}
\end{equation}
 which yields the eigenvalues of the rotational collective Hamiltonian
$H_{\mathrm{rot}}$ in Eq.~(\ref{eq:rotH}) as follows: 
\begin{eqnarray}
E_{(p,\, q),\, J} & = & \mathcal{M}_{\mathrm{cl}}
\;+\;\frac{1}{2}\left(\frac{1}{I_{1}}
\,-\,\frac{1}{I_{2}}\right)\,
J\,(J\,+\,1)
\cr
 &  & +\frac{1}{6I_{2}}\left(p^{2}
\,+\, q^{2}\,+\,3(p\,+\, q)\,+\, p\,
   q\right)\,-\,\frac{3}{8I_{2}}.
\label{eq:EpqJ}
\end{eqnarray}
 The allowed SU(3) baryon multiplets with zero triality are given as 
\begin{eqnarray}
(p,\, q) & = & (1,\,1)\;\;\rightarrow\;\; J=1/2\;\;(\mbox{octet}),\cr
(p,\, q) & = & (3,\,0)\;\;\rightarrow\;\; J=3/2\;\;(\mbox{decuplet}),\cr
(p,\, q) & = & (0,\,3)\;\;\rightarrow\;\;
J=1/2\;\;(\mbox{antidecuplet}).
\label{eq:baryonPQ}
\end{eqnarray}
 Thus, the mass splittings between the centers of the multiplets are
obtained as follows: 
\begin{eqnarray}
\Delta\overline{M}_{\mathbf{10-8}}
&=& E_{(3,0),\, J=3/2}\,-\, E_{(1,1),\, J=1/2}
=  \overline{M}_{\mathbf{10}}-\overline{M}_{\mathbf{8}} 
= \frac{3}{2\, I_{1}},
\cr
\Delta\overline{M}_{\mathbf{\overline{10}-8}}
&=& E_{(0,3),\, J=1/2}\,-\, E_{(1,1),\, J=1/2}
=  \overline{M}_{\mathbf{\overline{10}}}
-\overline{M}_{\mathbf{8}}  =\frac{3}{2\, I_{2}},
\cr
\Delta\overline{M}_{\mathbf{\overline{10}-10}}
&=& E_{(0,3),\, J=1/2}\,-\, E_{(3,0),\, J=3/2}
= \overline{M}_{\mathbf{\overline{10}}}
-\overline{M}_{\mathbf{10}}  =-\frac{3}{2\, I_{1}}+\frac{3}{2\,
  I_{2}},
\label{eq:endif}
\end{eqnarray}
 which shows that they arise from the rotational excitations. It is
well known and understood that in the $\chi\mathrm{SM}$ one cannot
calculate the absolute values of baryonic masses unless one incorporates
some non-relativistic corrections~\cite{Pobylitsa:1992bk}. We do
not do this in the present work and concentrate rather on the mass
splittings, which are all well defined. Thus, it is crucial to determine
the soliton moments of inertia $I_{1,2}$ uniquely. In all $\chi\mathrm{SM}$
calculations, $I_{1}$ turns out to be larger than $I_{2}$, which
leads to the consequence that the masses of the antidecuplet become
larger than those of the octet.

In order to determine the SU(3) baryon mass splittings, we now consider
the symmetry-breaking Hamiltonian $H_{\mathrm{sb}}$ in Eq.~(\ref{eq:sbH}).
The corrections due to the $\mathrm{SU(3)_{f}}$ and isospin symmetry
breaking effects are obtained perturbatively by calculating the matrix
elements of the $H_{\mathrm{sb}}$ between the diagonal baryon states
that are written as the SU(3) Wigner $D$ functions in representation
$\mathcal{R}$: 
\begin{eqnarray}
&&\langle A|\mathcal{R},\, B(Y\, T\, T_{3},\; Y^{\prime}\, J\,
J_{3})\rangle \cr
& = & \Psi_{(\mathcal{R}^{*}\,;\, Y^{\prime}\, J\,
  J_{3})}^{(\mathcal{R\,};\, Y\, T\, T_{3})}(A)
  =  \sqrt{\textrm{dim}(\mathcal{R})}\,(-)^{J_{3}
+Y^{\prime}/2}\, D_{(Y,\, T,\, T_{3})(-Y^{\prime},\,
J,\,-J_{3})}^{(\mathcal{R})*}(A),
\label{eq:Wigner}
\end{eqnarray}
 where $\mathcal{R}$ stands for the allowed irreducible representations
of the $\mathrm{SU(3)_{f}}$ group, i.e. 
$\mathcal{R}\,=\,8,\,10,\,\overline{10},\cdot\cdot\cdot$
and $Y,\, T,\, T_{3}$ are the corresponding hypercharge, isospin,
and its third component, respectively. The constraint of the right
hypercharge $Y^{\prime}\;=\;1$ selects a tower of allowed $\mathrm{SU(3)_{f}}$
representations: The lowest ones, that is, the baryon octet and decuplet,
coincide with those of the quark model. This has been considered as
a success of the collective quantization and as a sign of certain
duality between rigidly rotating heavy soliton and constituent quark
model. The third lowest representation is the antidecuplet which has
been considered as an artifact of the model and therefore disregarded
until the work of Diakonov \textit{et al}.~\cite{Diakonov:1997mm}.

Since the symmetry-breaking term of the collective Hamiltonian in
Eq.~(\ref{eq:sbH}) mixes different $\mathrm{SU(3)_{f}}$ representations,
the collective wave functions are no more in pure states but are given
as the following linear combinations \cite{Kim:1998gt}: 
\begin{align}
\left|B_{8}\right\rangle  
& =\left|8_{1/2},B\right\rangle 
\;+\; c_{\overline{10}}^{B}\left|\overline{10}_{1/2},B\right\rangle 
\;+\; c_{27}^{B}\left|27_{1/2},B\right\rangle ,
\cr
\left|B_{10}\right\rangle  
& =\left|10_{3/2},B\right\rangle 
\;+\; a_{27}^{B}\left|27_{3/2},B\right\rangle 
\;+\;
a_{35}^{B}\left|35_{3/2},B\right\rangle ,
\cr
\left|B_{\overline{10}}\right\rangle  
& =\left|\overline{10}_{1/2},B\right\rangle 
\;+\; d_{8}^{B}\left|8_{1/2},B\right\rangle 
\;+\; d_{27}^{B}\left|27_{1/2},B\right\rangle 
\;+\; d_{\overline{35}}^{B}\left|\overline{35}_{1/2},B\right\rangle ,
\label{eq:admix}
\end{align}
 where $\left|B_{\mathcal{R}}\right\rangle $ denotes the state which
reduces to the $\mathrm{SU(3)_{f}}$ representation $\mathcal{R}$
in the formal limit $m_{s}\rightarrow0$. Here, the spin indices $J_{3}$
have been suppressed. The $m_{s}$-dependent (through the linear $m_{s}$
dependence on $\alpha$, $\beta$ and $\gamma$) coefficients in
Eq.(\ref{eq:admix}) read: 
\begin{eqnarray}
c_{\overline{10}}^{B} & = &c_{\overline{10}}\left[\kern-0.5em 
\begin{array}{c}
\sqrt{5}\\
0\\
\sqrt{5}\\
0
\end{array}\kern-0.2em \right]\kern-0.2em ,\;
c_{27}^{B}=c_{27}\left[\kern-0.5em 
\begin{array}{c}
\sqrt{6}\\
3\\
2\\
\sqrt{6}
\end{array}\kern-0.2em \right]\kern-0.2em ,\;
a_{27}^{B}=a_{27}\left[\kern-0.5em 
\begin{array}{c}
\sqrt{15/2}\\
2\\
\sqrt{3/2}\\
0
\end{array}\kern-0.2em \right]\kern-0.2em ,\;
a_{35}^{B}=a_{35}\left[\kern-0.5em 
\begin{array}{c}
5/\sqrt{14}\\
2\sqrt{5/7}\\
3\sqrt{5/14}\\
2\sqrt{5/7}
\end{array}\kern-0.2em \right], \cr
d_{8}^{B} & = & d_{8}\left[
\begin{array}{c}
0\\
\sqrt{5}\\
\sqrt{5}\\
0
\end{array}\right],\;\;\; d_{27}^{B}=d_{27}\left[
\begin{array}{c}
0\\
\sqrt{3/10}\\
2/\sqrt{5}\\
\sqrt{3/2}
\end{array}\right],\;\;\;
d_{\overline{35}}^{B}=d_{\overline{35}}\left[
\begin{array}{c}
1/\sqrt{7}\\
3/(2\sqrt{14)}\\
1/\sqrt{7}\\
\sqrt{5/56}
\end{array}\right],
\label{eq:mix0}
\end{eqnarray}
 respectively in the basis $[N,\;\Lambda,\;\Sigma,\;\Xi]$,
 $[\Delta,\;\Sigma^{\ast},\;\Xi^{\ast},\;\Omega]$,
 $\left[\Theta^{+},\;
 N_{\overline{10}},\;\Sigma_{\overline{10}},\;\Xi_{\overline{10}}\right]$
 and analogous states in $\mathcal{R}=27,\;35,\;\overline{35}$, and 
\begin{eqnarray}
c_{\overline{10}} 
& = & -\frac{I_{2}}{15}\left(m_{s}-\hat{m}\right)
\left(\alpha+\frac{1}{2}\gamma\right),\;\;\;\;\; c_{27}
=-\frac{I_{2}}{25}\left(m_{s}-\hat{m}\right)\left(\alpha-\frac{1}{6}\gamma\right),
\cr
a_{27} 
& = &
-\frac{I_{2}}{8}\left(m_{s}-\hat{m}\right)\left(\alpha+\frac{5}{6}\gamma\right),
\;\;\;\;\;
a_{35}
=-\frac{I_{2}}{24}\left(m_{s}-\hat{m}\right)\left(\alpha-\frac{1}{2}\gamma\right),
\cr
d_{8} 
& = &
\frac{I_{2}}{15}\left(m_{s}-\hat{m}\right)\left(\alpha+\frac{1}{2}\gamma\right),
\;\;\;\;\;\;\;
d_{27}
\;=\;-\frac{I_{2}}{8}\left(m_{s}-\hat{m}\right)\left(\alpha-\frac{7}{6}\gamma\right),
\cr
d_{\overline{35}} & = &
-\frac{I_{2}}{4}\left(m_{s}-\hat{m}\right)\left(\alpha+\frac{1}{6}\gamma\right).
\label{eq:mix1}
\end{eqnarray}
 We will show later that these mixing coefficients are determined
uniquely in the present scheme.

\subsection{Electromagnetic corrections to SU(3) baryon masses}
The EM mass corrections to SU(3) baryon masses were already discussed
in Ref.~\cite{Yang:2010id}. However, since they consist of an essential
part of the present analysis, we will recapitulate them in this subsection.
The following baryonic two-point correlation functions of the EM current
will provide the EM mass corrections: 
\begin{equation}
M_{B}^{\mathrm{EM}}
\;=\;\frac{1}{2}\int d^{3}x\, d^{3}y\langle
B|T[J_{\mu}(\bm{x})J^{\mu}(\bm{y})]|B
\rangle D_{\gamma}(\bm{x},\bm{y})\;=\;\langle
B|\mathcal{O}^{\mathrm{EM}}|B\rangle,
\label{eq:corr}
\end{equation}
 where $J^{\mu}$ is defined as $J^{\mu}(x)\;=\;
 e\bar{\psi}(x)\gamma_{\mu}\hat{Q}\psi(x)$ with the electric charge
 $e$ and the quark charge operator $\hat{Q}$ defined as the
 Gell-Mann-Nishijima relation
 $\hat{Q}\;=\;\left(\lambda_{3}+\lambda_{8}/\sqrt{3}\right)/2$. The
 $D_{\gamma}$ denotes a static photon propagator which will be
 absorbed in parameters we will fit to experimental data. 
Using the fact that the EM current is taken as an octet operator,
we write the most general form of the $\mathcal{O}_{\mathrm{EM}}$
as a collective operator 
\begin{eqnarray}
\mathcal{O}^{\mathrm{EM}} 
& = & \alpha_{1}\sum_{i=1}^{3}D_{Qi}^{(8)}D_{Qi}^{(8)}
+\alpha_{2}\sum_{p=4}^{7}D_{Qp}^{(8)}D_{Qp}^{(8)}
+\alpha_{3}D_{Q8}^{(8)}D_{Q8}^{(8)},
\label{eq:emop}
\end{eqnarray}
 where $D_{Qa}^{(8)}=(D_{3a}^{(8)}+D_{8a}^{(8)}/\sqrt{3})/2$. The
 parameters $\alpha_{i}$ depend on specific dynamics of a $\chi$SM,
 which will be fitted to the empirical data of the EM mass
 differences. The product of two octet operators can be expanded in
 terms of irreducible operators
 $\mathbf{1}\oplus\mathbf{8_{s}}\oplus\mathbf{8_{a}}
\oplus\mathbf{10}\oplus\mathbf{\overline{10}}\oplus\mathbf{27}$.
 Note, however, that because of Bose symmetry we are left only with
 the singlet, the octet, and the eikosiheptaplet, which are all
 symmetric. We rewrite $\mathcal{O}^{\mathrm{EM}}$ in terms of a new
 set of parameters $c^{(n)}$ as follows: 
\begin{eqnarray}
\mathcal{O}^{\mathrm{EM}} 
& = & 
c^{(27)}\left(\sqrt{5}D_{\Sigma_{2}^{0}\Lambda_{27}}^{(27)}
+\sqrt{3}D_{\Sigma_{1}^{0}\Lambda_{27}}^{(27)}+D_{\Lambda_{27}\Lambda_{27}}^{(27)}\right)
\cr
 & + &
 c^{(8)}\left(\sqrt{3}D_{\Sigma^{0}\Lambda}^{(8)}+D_{\Lambda\Lambda}^{(8)}\right)
\;+\; c^{(1)}D_{\Lambda\Lambda}^{(1)}.
\label{eq:emop3}
\end{eqnarray}
 The last term in Eq.(\ref{eq:emop3}) does not contribute to the
mass splittings. The EM mass can be obtained by sandwiching the collective
operator $\mathcal{O}_{\mathrm{EM}}$ in Eq.(\ref{eq:emop}) between
the baryon states. The corresponding results can be written for the
baryon octet 
\begin{eqnarray}
M_{N}^{\mathrm{EM}} & = & 
\frac{1}{5}\left(c^{(8)}
+\frac{4}{9}c^{(27)}\right)T_{3}
+\frac{3}{5}\left(c^{(8)}
+\frac{2}{27}c^{(27)}\right)\left(T_{3}^{2}
+\frac{1}{4}\right)+c^{(1)},
\cr
M_{\Lambda}^{\mathrm{EM}} & = & 
\frac{1}{10}\left(c^{(8)}
-\frac{2}{3}c^{(27)}\right)+c^{(1)},
\cr
M_{\Sigma}^{\mathrm{EM}} & = & 
\frac{1}{2}c^{(8)}\, T_{3}
+\frac{2}{9}c^{(27)}\, T_{3}^{2}
-\frac{1}{10}\left(c^{(8)}
+\frac{14}{9}c^{(27)}\right)+c^{(1)},
\cr
M_{\Xi}^{\mathrm{EM}} & = & 
\frac{4}{5}\left(c^{(8)}
-\frac{1}{9}c^{(27)}\right)T_{3}
-\frac{2}{5}\left(c^{(8)}
-\frac{1}{9}c^{(27)}\right)\left(T_{3}^{2}
+\frac{1}{4}\right)+c^{(1)},
\label{eq:emforoc}
\end{eqnarray}
 and for the baryon decuplet 
\begin{eqnarray}
M_{\Delta}^{\mathrm{EM}} 
& = & \frac{1}{4}\left(c^{(8)}
+\frac{8}{63}c^{(27)}\right)T_{3}
+\frac{5}{63}c^{(27)}\,
T_{3}^{2}+\frac{1}{8}\left(c^{(8)}
-\frac{2}{3}c^{(27)}\right)+c^{(1)},
\cr
M_{\Sigma^{\ast}}^{\mathrm{EM}} 
& = &
\frac{1}{4}\left(c^{(8)}
-\frac{4}{21}c^{(27)}\right)T_{3}
+\frac{5}{63}c^{(27)}\,\left(T_{3}^{2}-1\right)+c^{(1)},
\cr
M_{\Xi^{\ast}}^{\mathrm{EM}} 
& = &
\frac{1}{4}\left(c^{(8)}
-\frac{32}{63}c^{(27)}\right)T_{3}
-\frac{1}{4}\left(c^{(8)}
+\frac{8}{63}c^{(27)}\right)\left(T_{3}^{2}
+\frac{1}{4}\right)+c^{(1)},
\cr
M_{\Omega^{-}}^{\mathrm{EM}} 
& = &
-\frac{1}{4}\left(c^{(8)}
-\frac{4}{21}c^{(27)}\right)+c^{(1)},
\label{eq:emfordec}
\end{eqnarray}
 respectively. Since the center of baryon masses can absorb the
 singlet contributions to the EM masses with $c^{(1)}$, we can safely
 neglect them for EM mass differences. Moreover, they are not
 pertinent to the EM mass differences in which they are canceled
 out. Therefore, the expressions of the EM mass differences of SU(3)
 baryons have only two unknown parameters, i.e. $c^{(8)}$ and
 $c^{(27)}$. 
As shown in Eqs.(\ref{eq:emforoc}, \ref{eq:emfordec}), they are
expressed in terms of the isospin third component $T_{3}$, its square
$T_{3}^{2}$, and the constant terms arising from the hypercharge.
Note that Eqs.(\ref{eq:emforoc}, \ref{eq:emfordec}) in general can
be rewritten in terms of the electric charge $Q$ and its square $Q^{2}$.
Moreover, it turns out that Eqs.(\ref{eq:emforoc}, \ref{eq:emfordec})
have the same structures as the Weinberg-Treiman mass formula
 $M(T_{3})\;=\; aT_{3}^{2}+bT_{3}+c$~\cite{Weinberg:1959zzb} 
expressed in terms of three free parameters $a$, $b$, and $c$.

In Refs.~\cite{Praszalowicz:1992gn,Blotz:1994pc}, the Dashen
ansatz~\cite{Dashen} was used for the EM mass splittings of the SU(3)
baryons, which shows $Q^{2}$ proportionality ($(\Delta
M_{B})_{\mathrm{EM}}\sim Q_{B}^{2}M_{B}$). However, this Ansatz was
originally used for the squares of SU(3) pseudoscalar meson masses and
is valid only in the chiral limit. In fact,
Ref.~\cite{Praszalowicz:1992gn} employed the Dashen Ansatz and fixed
the unknown free parameter appearing in this Ansatz, using the result
of $(\Sigma^{-}+\Sigma^{+}-2\Sigma^{0})$ derived in
Ref.~\cite{Gasser:1982ap}. However, this Ansatz does not determine the
sign of the EM mass splittings. 
It is straightforward to obtain the EM mass differences for the baryon
octet from Eq.(\ref{eq:emforoc}) 
\begin{eqnarray}
(M_{p}-M_{n})_{\mathrm{EM}} 
& = & \frac{1}{5}\left(c^{(8)}+\frac{4}{9}c^{(27)}\right),
\;\;\;\;\;(M_{\Sigma^{+}}-M_{\Sigma^{-}})_{\mathrm{EM}}\;=\; c^{(8)},\cr
(M_{\Xi^{0}}-M_{\Xi^{-}})_{\mathrm{EM}} & = &
\frac{4}{5}\left(c^{(8)}-\frac{1}{9}c^{(27)}\right).
\label{eq:diff_EM}
\end{eqnarray}
 Using Eq.(\ref{eq:diff_EM}), we immediately obtain the following
mass formula
$c^{(8)}=(M_{p}-M_{n})_{\mathrm{EM}}+(M_{\Xi^{0}}-M_{\Xi^{-}})_{\mathrm{EM}}
=(M_{\Sigma^{+}}-M_{\Sigma^{-}})_{\mathrm{EM}}$.
This is just the well-known Coleman-Glashow mass formula~\cite{Coleman:1961jn}.
Although these formulae indicate that these three mass differences
are dependent on each other, one can adjust the values of the parameters
$c^{(8)}$ and $c^{(27)}$ by the method of least squares. In order
to determine the parameters $c^{(8)}$ and $c^{(27)}$, we will first
use the empirical data estimated in Ref.~\cite{Gasser:1982ap}. Using
these empirical and experimental data, we can determine the values
of the parameters $c^{(8)}$ and $c^{(27)}$ as follows 
\begin{equation}
c^{(8)}\;=\;-0.15\pm0.23,\;\;\; c^{(27)}\;=\;8.62\pm2.39,
\label{eq:c8c27}
\end{equation}
 in units of MeV~\cite{Yang:2010id}.

\subsection{Baryon octet\label{sub:Baryon-octet}}

The effects of $\mathrm{SU(3)_{f}}$ and isospin symmetry breakings
being taken into account, the mass formulae of the octet are obtained 
as follows: 
\begin{eqnarray}
M_{N} & = & \overline{M}_{\mathbf{8}}+c^{(1)}
+\frac{1}{5}\left(c^{(8)}
+\frac{4}{9}c^{(27)}\right)T_{3}
+\frac{3}{5}\left(c^{(8)}
+\frac{2}{27}c^{(27)}\right)\left(T_{3}^{2}+\frac{1}{4}\right)
\cr
&  & -\left(m_{d}-m_{u}\right)\left(\delta_{1}-\delta_{2}\right)T_{3}
-\left(m_{s}-\hat{m}\right)\left(\delta_{1}+\delta_{2}\right),\cr
M_{\Lambda} 
& = &
\overline{M}_{\mathbf{8}}+c^{(1)}+\frac{1}{10}\left(c^{(8)}
-\frac{2}{3}c^{(27)}\right)-\left(m_{s}-\hat{m}\right)\delta_{2},
\cr
M_{\Sigma} 
& = &
\overline{M}_{\mathbf{8}}+c^{(1)}+\frac{1}{2}c^{(8)}\, T_{3}
+\frac{2}{9}c^{(27)}\, T_{3}^{2}-\frac{1}{10}\left(c^{(8)}
+\frac{14}{9}c^{(27)}\right)\cr
 &  &
 -\left(m_{d}-m_{u}\right)\left(\delta_{1}
+\frac{1}{2}\delta_{2}\right)T_{3}
+\left(m_{s}-\hat{m}\right)\delta_{2},\cr
M_{\Xi} & = & \overline{M}_{\mathbf{8}}+c^{(1)}
+\frac{4}{5}\left(c^{(8)}-\frac{1}{9}c^{(27)}\right)T_{3}
-\frac{2}{5}\left(c^{(8)}-\frac{1}{9}c^{(27)}\right)\left(T_{3}^{2}+\frac{1}{4}\right)
\cr
 &  & -\left(m_{d}-m_{u}\right)\left(\delta_{1}
+2\delta_{2}\right)T_{3}+\left(m_{s}-\hat{m}\right)\delta_{1},
\label{eq:octM}
\end{eqnarray}
 where $\delta_{1}$ and $\delta_{2}$ are defined as 
\begin{eqnarray}
\delta_{1} 
& = & -\frac{1}{5}\alpha\;-\;\beta\;+\;\frac{1}{5}\gamma,
\cr
\delta_{2} 
& = & -\frac{1}{10}\alpha\;-\;\frac{3}{20}\gamma.
\label{eq:d1d2}
\end{eqnarray}
 The center mass $\overline{M}_{\mathbf{8}}$ for the baryon octet
from Eq. (\ref{eq:octM}) is found to be 
\begin{eqnarray}
\overline{M}_{\mathbf{8}} 
& = &
\frac{1}{3}\overline{M}_{N}+\frac{1}{4}\overline{M}_{\Sigma}
+\frac{1}{12}M_{\Sigma^{0}}
+\frac{1}{3}\overline{M}_{\Xi}-c^{(1)},
\label{eq:8center}
\end{eqnarray}
 where $\overline{M}_{B}$ indicates the mean values of the masses
in the corresponding isospin multiplets, for example, 
$\overline{M}_{N}\;=\;\left(M_{p}+M_{n}\right)/2$.
Note that the $\overline{M}_{\mathbf{8}}$ is expressed in terms of
octet masses.

\subsection{Baryon decuplet\label{sub:Baryon-decuplet}}

Similarly, we can derive the masses of the baryon decuplet with the
center of the decuplet $\overline{M}_{\mathbf{10}}$: 
\begin{eqnarray}
M_{\Delta} 
& = & \overline{M}_{\mathbf{10}}+c^{(1)}+\frac{1}{4}\left(c^{(8)}
+\frac{8}{63}c^{(27)}\right)T_{3}+\frac{5}{63}c^{(27)}\, T_{3}^{2}
+\frac{1}{8}\left(c^{(8)}-\frac{2}{3}c^{(27)}\right)\cr
 &  &
 -\left(m_{d}-m_{u}\right)\left(\delta_{1}-\frac{3}{4}\delta_{2}\right)T_{3}
-\left(m_{s}-\hat{m}\right)\left(\delta_{1}-\frac{3}{4}\delta_{2}\right),
\cr
M_{\Sigma^{\ast}} 
& = & \overline{M}_{\mathbf{10}}+c^{(1)}+\frac{1}{4}\left(c^{(8)}
-\frac{4}{21}c^{(27)}\right)T_{3}+\frac{5}{63}c^{(27)}\,\left(T_{3}^{2}-1\right)
\cr
 &  &
 -\left(m_{d}-m_{u}\right)\left(\delta_{1}-\frac{3}{4}\delta_{2}\right)T_{3},
\cr
M_{\Xi^{\ast}} 
& = & \overline{M}_{\mathbf{10}}+c^{(1)}+\frac{1}{4}\left(c^{(8)}
-\frac{32}{63}c^{(27)}\right)T_{3}-\frac{1}{4}\left(c^{(8)}
+\frac{8}{63}c^{(27)}\right)\left(T_{3}^{2}+\frac{1}{4}\right)\cr
 &  &
 -\left(m_{d}-m_{u}\right)\left(\delta_{1}-\frac{3}{4}\delta_{2}\right)T_{3}
+\left(m_{s}-\hat{m}\right)\left(\delta_{1}-\frac{3}{4}\delta_{2}\right),\cr
M_{\Omega^{-}} 
& = &
\overline{M}_{\mathbf{10}}+c^{(1)}
-\frac{1}{4}\left(c^{(8)}-\frac{4}{21}c^{(27)}\right)
+2\left(m_{s}-\hat{m}\right)\left(\delta_{1}-\frac{3}{4}\delta_{2}\right).
\label{eq:decM}
\end{eqnarray}
 As in the case of the baryon octet, the center of mass splittings
$\overline{M}_{\mathbf{10}}$ of the baryon decuplet can be expressed
as 
\begin{eqnarray}
\overline{M}_{\mathbf{10}} 
& = &
\frac{3}{2}\overline{M}_{\Sigma^{\ast}}
-\frac{1}{2}M_{\Sigma^{\ast0}}-c^{(1)}.
\label{eq:1st10center}
\end{eqnarray}
 Making use of Eq. (\ref{eq:decM}), we are able to obtain various
mass relations among the decuplet baryons as follows: 
\begin{eqnarray}
M_{\Delta^{\text{++}}}-M_{\Delta^{+}} 
& = & \left(M_{\Xi^{\ast0}}-M_{\Xi^{\ast-}}\right)+2\Delta
M_{\Sigma^{\ast}},
\cr
M_{\Delta^{+}}-M_{\Delta^{0}} 
& = & M_{\Sigma^{\ast+}}-M_{\Sigma^{\ast0}}
\;\;=\;\;\left(M_{\Xi^{\ast0}}-M_{\Xi^{\ast-}}\right)\;+\;\Delta
M_{\Sigma^{\ast}},
\cr
M_{\Delta^{0}}-M_{\Delta^{-}} 
& = & M_{\Sigma^{\ast0}}-M_{\Sigma^{\ast-}}
\;\;=\;\;\left(M_{\Xi^{\ast0}}-M_{\Xi^{\ast-}}\right)
\label{eq:10rel1}
\end{eqnarray}
 for $\Delta T_{3}\;=\;1$, and 
\begin{eqnarray}
M_{\Delta^{\text{++}}}-M_{\Delta^{0}} 
& = & 2\left(M_{\Xi^{\ast0}}-M_{\Xi^{\ast-}}\right)+3\Delta
M_{\Sigma^{\ast}},
\cr
M_{\Delta^{+}}-M_{\Delta^{-}} 
& = & M_{\Sigma^{\ast+}}-M_{\Sigma^{\ast-}}
\;\;=\;\;2\left(M_{\Xi^{\ast0}}-M_{\Xi^{\ast-}}\right)\;+\;\Delta
M_{\Sigma^{\ast}}
\label{eq:10rel2}
\end{eqnarray}
 for $\Delta T_{3}\;=\;2$. In the case of $\Delta T_{3}\;\;=\;\;3$,
we get 
\begin{eqnarray}
M_{\Delta^{\text{++}}}-M_{\Delta^{-}} 
& = & 3\left(M_{\Xi^{\ast0}}-M_{\Xi^{\ast-}}\right)+3\Delta
M_{\Sigma^{\ast}}.
\label{eq:10rel3}
\end{eqnarray}
 In addition, we derive one more mass formula as follows: 
\begin{equation}
2M_{\Xi^{*-}}-M_{\Sigma^{*-}}\;=\; M_{\Omega^{-}}\,.
\label{eq:10rel4}
\end{equation}
 Equations~(\ref{eq:10rel1})-(\ref{eq:10rel4}) generalize the
 Gell-Mann-Okubo mass formulae. While the experimental data for the
 $\Delta$ isobars are not enough to judge the above-given mass
 relations, all other relations apart from the $\Delta$ isobars are
 all in good agreement with the data. If we turn off isospin symmetry
 breaking, Eq.~(\ref{eq:decM}) is reduced to the Gell-Mann-Okubo mass
 formula for the decuplet~\cite{GellMann:1962xb,Okubo:1961jc} as
 follows:  
\begin{eqnarray}
\left(\overline{M}_{\Delta}-\overline{M}_{\Sigma^{\ast}}\right) 
& = &
\left(\overline{M}_{\Sigma^{\ast}}-\overline{M}_{\Xi^{\ast}}\right)
\;=\;\left(\overline{M}_{\Xi^{\ast}}-\overline{M}_{\Omega^{-}}\right),
\cr
\overline{M}_{\Delta}-\overline{M}_{\Omega^{-}} 
& = &
3\left(\overline{M}_{\Sigma^{\ast}}-\overline{M}_{\Xi^{\ast}}\right).
\label{eq:10_equalspacing}
\end{eqnarray}

In a $\chi$SM, it is also possible to connect the mass splittings
of the baryon octet to those of the decuplet~\cite{Guadagnini:1983uv}.
Since we include both flavor SU(3) and isospin symmetry breakings,
we can derive the following formulae 
\begin{eqnarray}
 &  & 2(M_{p}+M_{\Xi^{0}})
+3(-M_{\Sigma^{*-}}+2M_{\Xi^{*-}})
=3M_{\Lambda}+2M_{\Sigma^{+}}-M_{\Sigma^{0}}+3M_{\Omega^{-}},
\cr
 &  & 2(M_{n}+M_{\Xi^{-}})
+3(-M_{\Sigma^{*-}}+2M_{\Xi^{*-}})
=3M_{\Lambda}+2M_{\Sigma^{-}}-M_{\Sigma^{0}}+3M_{\Omega^{-}}\,.
\label{eq:guag}
\end{eqnarray}
 These are the generalization of the Guadagnini mass formula and are
deviated from the experimental data by about 4 MeV only, which is
remarkable. The basically same formula was obtained in
Ref.~\cite{Morpurgo:1991if}. Using Eq.~(\ref{eq:10rel4}) and turning
off the effects of isospin symmetry breaking, we reproduce the
Guadagnini formula~\cite{Guadagnini:1983uv} 
\begin{eqnarray}
8\left(\overline{M}_{N}+\overline{M}_{\Xi^{\ast}}\right)+3\,\overline{M}_{\Sigma} 
& = & 11\,\overline{M}_{\Lambda}+8\,\overline{M}_{\Sigma^{\ast}}\,.
\label{eq:genGua}
\end{eqnarray}

\subsection{Baryon antidecuplet \label{sub:Baryon-antidecuplet}}
We now come to the expressions of the mass splittings of the baryon
antidecuplet. The masses of the antidecuplet are expressed as 
\begin{eqnarray}
M_{\Theta^{+}} 
& = & \overline{M}_{\mathbf{\overline{10}}}
+c^{(1)}+\frac{1}{4}\left(c^{(8)}
-\frac{4}{21}c^{(27)}\right)
-2\left(m_{s}
-\hat{m}\right)\delta_{3},
\cr
M_{N^{\ast}} 
& = & \overline{M}_{\mathbf{\overline{10}}}
+c^{(1)}
+\frac{1}{4}\left(c^{(8)}
-\frac{32}{63}c^{(27)}\right)T_{3}
+\frac{1}{4}\left(c^{(8)}
+\frac{8}{63}c^{(27)}\right)\left(T_{3}^{2}
+\frac{1}{4}\right)
\cr
 &  & -\left(m_{d}-m_{u}\right)\delta_{3}T_{3}
-\left(m_{s}-\hat{m}\right)\delta_{3},
\cr
M_{\Sigma_{\overline{10}}} 
& = & \overline{M}_{\mathbf{\overline{10}}}
+c^{(1)}+\frac{1}{4}\left(c^{(8)}
-\frac{4}{21}c^{(27)}\right)T_{3}\cr
&&
-\frac{5}{63}c^{(27)}\left(T_{3}^{2}-1\right)
-\left(m_{d}-m_{u}\right)\delta_{3}T_{3},
\cr
M_{\Xi_{3/2}^{+}} 
& = & \overline{M}_{\mathbf{\overline{10}}}
+c^{(1)}+\frac{1}{4}\left(c^{(8)}
+\frac{8}{63}c^{(27)}\right)T_{3}
-\frac{5}{63}c^{(27)}T_{3}^{2}
-\frac{1}{8}\left(c^{(8)}-\frac{2}{3}c^{(27)}\right)
\cr
 &  &
 -\left(m_{d}-m_{u}\right)\delta_{3}T_{3}
+\left(m_{s}-\hat{m}\right)\delta_{3},
\label{eq:antidecM}
\end{eqnarray}
 where the center of the mass splittings of the baryon antidecuplet
is given as 
\begin{equation}
\overline{M}_{\mathbf{\overline{10}}}
\;=\;\frac{3}{2}\overline{M}_{\Sigma_{\overline{10}}}
-\frac{1}{2}M_{\Sigma_{\overline{10}}^{0}}-c^{(1)}
\label{eq:antiMo}
\end{equation}
and $\delta_{3}$ is defined as 
\begin{equation}
\delta_{3}
\;=\;{\displaystyle
  -\frac{1}{8}\alpha
\;-\;\beta
\;+\;\frac{1}{16}\gamma}.
\label{eq:delta3}
\end{equation}
 Note that while the masses of the baryon octet and decuplet contain
$\delta_{1}$ and $\delta_{2}$, those of the antidecuplet include
$\delta_{3}$.

Using Eq. (\ref{eq:antidecM}), we can derive similar mass relations
to Eqs.~(\ref{eq:octM}, \ref{eq:decM}) as follows: 
\begin{eqnarray}
M_{\Xi_{3/2}^{+}}-M_{\Xi_{3/2}^{0}} 
& = & M_{\Sigma_{\overline{10}}^{+}}
-M_{\Sigma_{\overline{10}}^{0}}
\;\;=\;\;\left(M_{p^{\ast}}-M_{n^{\ast}}\right)
\cr
M_{\Xi_{3/2}^{0}}-M_{\Xi_{3/2}^{-}} 
& = & M_{\Sigma_{\overline{10}}^{0}}
-M_{\Sigma_{\overline{10}}^{-}}
\;\;=\;\;\left(M_{p^{\ast}}-M_{n^{\ast}}\right)
\;-\;\Delta M_{\Sigma_{\overline{10}}},
\cr
M_{\Xi_{3/2}^{-}}-M_{\Xi_{3/2}^{--}} 
& = & \left(M_{p^{\ast}}-M_{n^{\ast}}\right)-2\Delta
M_{\Sigma_{\overline{10}}},
\label{eq:10barrel1}
\end{eqnarray}
 for $\Delta T_{3}\;=\;1$, and 
\begin{eqnarray}
M_{\Xi_{3/2}^{+}}-M_{\Xi_{3/2}^{-}} 
& = & M_{\Sigma_{\overline{10}}^{+}}-M_{\Sigma_{\overline{10}}^{-}}
\;\;=\;\;2\left(M_{p^{\ast}}-M_{n^{\ast}}\right)
\;-\;\Delta M_{\Sigma_{\overline{10}}},
\cr
M_{\Xi_{3/2}^{0}}-M_{\Xi_{3/2}^{--}} 
& = & 2\left(M_{p^{\ast}}-M_{n^{\ast}}\right)-3\Delta
M_{\Sigma_{\overline{10}}},
\label{eq:10barrel2}
\end{eqnarray}
 for $\Delta T_{3}\;=\;2$. In the case of $\Delta T_{3}\;\;=\;\;3$,
we get 
\begin{eqnarray}
M_{\Xi_{3/2}^{+}}-M_{\Xi_{3/2}^{--}} 
& = & 3\left(M_{p^{\ast}}-M_{n^{\ast}}\right)
-3\Delta M_{\Sigma_{\overline{10}}},
\label{eq:10barrel3}
\end{eqnarray}
 where $\Delta M_{\Sigma_{\overline{10}}}
\;\;=\;\; M_{\Sigma_{\overline{10}}^{+}}
+M_{\Sigma_{\overline{10}}^{-}}-2M_{\Sigma_{\overline{10}}^{0}}$.
The effects of isospin symmetry breaking being switched off, the mass
formula of Ref.~\cite{Diakonov:1997mm} is reproduced as 
\begin{eqnarray}
\overline{M}_{\Theta^{+}}-\overline{M}_{\Xi_{3/2}} 
& = & 3\left(\overline{M}_{N^{\ast}}
-\overline{M}_{\Sigma_{\overline{10}}}\right).
\label{eq:new10rel2}
\end{eqnarray}
 In addition, we obtain the new mass relations between the baryon
octet and antidecuplet: 
\begin{equation}
\left(2\overline{M}_{N^{\ast}}\,+\,2\overline{M}_{N}
\;-\;3M_{\Lambda}\;-\;\overline{M}_{\Sigma_{\overline{10}}}\right)
\;=\;\left(\overline{M}_{\Sigma}\;-\;2\overline{M}_{\Xi}
\;+\; M_{\Theta^{+}}\right)
\label{eq:8rel10bar}
\end{equation}
 and 
\begin{equation}
3\left(\overline{M}_{\Sigma^{\ast}}
+M_{\Omega^{-}}\right)
+2\left(M_{\Theta^{+}}
+2\overline{M}_{\Xi_{3/2}}\right)
\;=\;6\left(\overline{M}_{\Xi^{\ast}}
+\overline{M}_{\Sigma_{\overline{10}}}\right)
\label{eq:10-10bar-off}
\end{equation}
 between the baryon decuplet and antidecuplet. We also get the mass
relation among the baryon octet, decuplet, and antidecuplet 
\begin{eqnarray}
 &  & \left(11M_{\Lambda}+5\overline{M}_{\Sigma^{\ast}}
+6\overline{M}_{\Sigma_{\overline{10}}}\right)
\cr
 & = & 3\left(M_{\Omega^{-}}+\overline{M}_{\Sigma}\right)
+2\left(M_{\Theta^{+}}+\overline{M}_{\Xi^{\ast}}\right)
+4\left(2\overline{M}_{N}+\overline{M}_{\Xi_{3/2}}\right).
\label{eq:8-10-10bar-off}
\end{eqnarray}

\section{Results and Discussion}

By the least squared method, the model parameters can be adjusted
from the studies of mass splittings with the experimental mass values
of the baryon octet, $\Omega^{-}(1672)$, and $\Theta^{+}(1524)$
taken as inputs. The effcts of SU(3) flavor and isospin symmetry breakings
are obtained from the baryon octet mass splittings. The masses of
$\Omega^{-}(1672)$ and $\Theta^{+}(1524)$ are taken for determination
of the mass-splitting centers of the baryon decuplet and antidecuplet,
respectively. The centers of mass splittings also can be expressed
in terms of the model parameters, the moments of inertia of soliton
$I_{1}$ and $I_{2}$ in Eq. (\ref{eq:endif}). The Eq. (\ref{eq:octM})
yields the ratio of the current light quark masses as follows: 
\begin{eqnarray}
R\; & = & \;\frac{m_{s}-\hat{m}}{m_{d}-m_{u}}
\cr
 & = & \frac{M_{p}-M_{\Sigma^{+}}+M_{\Sigma^{0}}
-M_{\Xi^{-}}}{2\left(M_{\Sigma^{+}}-M_{\Sigma^{0}}\right)},
\label{eq:Rvalue}
\end{eqnarray}
 which yields 
\begin{equation}
R\;=\;58.1\pm1.3,
\label{eq:Rval}
\end{equation}
Note that in Ref.~\cite{Gasser:1982ap} it is given as
$R\;=\;43.5\pm2.2$, which implies that $R$ in this work is comparable
to that of Ref.~\cite{Gasser:1982ap}. Using the experimental data for
the baryon octet, $\Omega$ and $\Theta^+$ with the value of $R$ in 
Eq.~(\ref{eq:Rval}), we can determine the mass parameters   
\begin{eqnarray}
\left(m_{d}-m_{u}\right)\alpha 
& = & -4.390\pm0.004,\;\;\;\;\;
\left(m_{s}-\hat{m}\right)\alpha\;=\;-255.029\pm5.821,
\cr
\left(m_{d}-m_{u}\right)\beta 
& = & -2.411\pm0.001,\;\;\;\;\;\;\;
\left(m_{s}-\hat{m}\right)\beta\;=\;-140.040\pm3.195,
\cr
\left(m_{d}-m_{u}\right)\gamma 
& = & -1.740\pm0.006,\;\;\;\;\;\;\;
\left(m_{s}-\hat{m}\right)\gamma\;=\;-101.081\pm2.332,
\label{eq:Nabrfinal}
\end{eqnarray}
in units of MeV. With these parameters, we find the moments of
inertia $I_1$ and $I_2$. Though $I_1$ and $I_2$ could be determined
by using Eqs.(\ref{eq:endif}) and the centers of mass splittings
defined in 
Eqs.~(\ref{eq:8center},\ref{eq:1st10center},\ref{eq:antiMo}), this
is impractical. We do not know experimentally
$M_{\Sigma_{\overline{10}}}$. Thus, in order to determine $I_1$ and 
$I_2$ we will rather use the mass formulae for $\Omega^-$ and
$\Theta^+$ in Eqs.(\ref{eq:decM}, \ref{eq:antidecM}), combining
them with the mass parameters determined in
Eq.(\ref{eq:Nabrfinal}). The obtained values of $I_1$ and $I_2$ are
listed in Table~\ref{tab:comp}. 

\begin{table}
\caption{The comparison of the present results of important parameters with
those of other works. The masses of the baryon antidecuplet members
used as input are listed in the second row.  Note that the results listed in the final
column do not contain the isospin symmetry breaking effects, so that
they are slightly different from those of Eq.(\ref{eq:Nabrfinal}).}

\begin{tabular}{c||c|c|c|c|c}
\hline 
\multicolumn{2}{c|}{} 
& Diakonov et al.~\cite{Diakonov:1997mm}  
& \multicolumn{1}{c|}{Ellis et al.~\cite{Ellis:2004uz}} 
& $\chi\mathrm{QSM}$\cite{Ledwig:2008rw}  
& This work \tabularnewline
\hline 
\multicolumn{2}{c|}{Input } 
& $N^{\ast}(1710\;\mathrm{MeV})$  
& $\Theta^{+}(1539\pm2\;\mathrm{MeV)}$  
& $\cdots$  
& $\Theta^{+}(1524\pm5\;\mathrm{MeV})$ 
\tabularnewline
\multicolumn{2}{c|}{masses} 
&  
& $\Xi^{--} _{3/2} (1862\pm2\;\mathrm{MeV})$  
& $\cdots$  
& \tabularnewline
\hline 
\multicolumn{2}{c|}{$\Sigma_{\pi N}$} 
& $45\;\mathrm{MeV}^{\star}$  
& $73\;\mathrm{MeV}$
& $41\;\mathrm{MeV}$  
& $36.4\pm3.9\;\mathrm{MeV}$ 
\tabularnewline
\hline 
\multicolumn{2}{c|}{$I_{1}$} 
& $1.29\;\mathrm{fm}$  
& $1.27\;\mathrm{fm}$  
& $1.06\;\mathrm{fm}$  
& $1.230\pm0.002\;\mathrm{fm}$ 
\tabularnewline
\multicolumn{2}{c|}{$I_{2}$} 
& $0.4\;\mathrm{fm}$  
& $0.49\;\mathrm{fm}$  
& $0.48\;\mathrm{fm}$  
& $0.420\pm0.006\;\mathrm{fm}$ 
\tabularnewline
\multicolumn{2}{c|}{$m_{s}\alpha$} 
& $-218\;\mathrm{MeV}$  & $-605\;\mathrm{MeV}$  
& $-197\;\mathrm{MeV}$  & $-262.9\pm5.9\;\mathrm{MeV}$ 
\tabularnewline
\multicolumn{2}{c|}{$m_{s}\beta$} 
& $-156\;\mathrm{MeV}$  & $-23\;\mathrm{MeV}$  
& $-94\;\mathrm{MeV}$  & $-144.3\pm3.2\;\mathrm{MeV}$ 
\tabularnewline
\multicolumn{2}{c|}{$m_{s}\gamma$} 
& $-107\;\mathrm{MeV}$  & $152\;\mathrm{MeV}$  
& $-53\;\mathrm{MeV}$  & $-104.2\pm2.4\;\mathrm{MeV}$ 
\tabularnewline
\multicolumn{2}{c|}{$c_{\overline{10}}$} & $0.084$  & $0.088$  &
$0.037$  & $0.0434\pm0.0006$ 
\tabularnewline
\hline 
\end{tabular}\label{tab:comp} 
\end{table}
The mixing coefficients defined in Eq.~(\ref{eq:mix0}) are fixed to be  
\begin{eqnarray}
c_{\overline{10}} & = & 0.0434\pm0.0006,
\;\;\;\;\;\;\; c_{27}=0.0203\pm0.0003,
\cr
a_{27} & = & 0.0903\pm0.0013,\;\;\;\;\;\;\; a_{35}=0.0181\pm0.0003,
\cr
d_{8} & = & -0.0434\pm0.0006,\;\;\;\; d_{27}=0.0365\pm0.0005,
\cr
d_{\overline{35}} & = & 0.1447\pm0.0021.
\label{eq:Nmix1}
\end{eqnarray}
 Employing the value of the ratio 
$\left(m_{d}-m_{u}\right)/\left(m_{d}+m_{u}\right)\;=\;0.28\pm0.03$
\cite{Gasser:1982ap}, we find the pion-nucleon sigma term
unambiguosly: 
\begin{equation}
\Sigma_{\pi N}\;=\;(36.4\pm3.9)\,\mathrm{MeV}.
\label{eq:Nsigma}
\end{equation}
 In Ref.~\cite{Diakonov:1997mm}, $\Sigma_{\pi N}=45\,\mathrm{MeV}$ was
 used~\cite{Gasser:1990ce}, while Ref.~\cite{Ellis:2004uz} predicted
 $\Sigma_{\pi N}=73\,\mathrm{MeV}$ in studying the baryon
 antidecuplet~\cite{Pavan:2001wz,Schweitzer:2003fg}. In fact, larger
 values of the $\Sigma_{\pi N}$ are predicted to describe the mass
 splitting in the baryon
 antidecuplet~\cite{Schweitzer:2003fg}. Indeed, the larger value of
 the $\Sigma_{\pi N}$ reduces the antidecuplet splitting
 noticeably~\cite{Diakonov:2003jj}. In Ref.~\cite{Schweitzer:2003fg},
 the $\Sigma_{\pi N}$ has been extracted by using the $\Theta^{+}$ and
 $\Xi_{3/2}$ masses, based on the $\chi$QSM: $\Sigma_{\pi
 N}=(74\pm12)\,\mathrm{MeV}$. However, the present result of
 $\Sigma_{\pi N}$ remains rather smaller than the previous analyses
 based on the masses of the baryon antidecuplet.

In Table~\ref{tab:comp}, we compare the present results of the
important parameters with those of other works. Note that
Ref.~\cite{Diakonov:1997mm} use the $\pi N$ sigma term
as input, while in the present work we are able to predict its value,
since we have considered the effects of isospin symmetry breaking. The
predicted value of $\Sigma_{\pi N}$ in this work is different from
that of the recent calculation in the $\chi$QSM, which is listed in
the fourth column. The results of moments of inertia $I_1$ and $I_{2}$
are comparable to those of 
Refs.~\cite{Diakonov:1997mm,Ellis:2004uz}. However, the important 
parameters $\alpha$, $\beta$, and $\gamma$ turn out to be rather
different. Even the sign of $\gamma$ in Ref.~\cite{Ellis:2004uz} is
different from the present result. The result of the mixing parameter
$c_{\overline{10}}$ turns out to be almost two times less than those
of Refs.~\cite{Diakonov:1997mm,Ellis:2004uz} while it is comparable to
that of the $\chi$QSM. This parameter is of great importance to
determine the coupling constants for the $K^{*}N\Theta^{+}$
vertex~\cite{Ledwig:1900ri}. For example, if we use the values of
$c_{\overline{10}}$ in Refs.~\cite{Diakonov:1997mm,Ellis:2004uz}, the
vector coupling constant
$g_{K^{*}n\Theta}$($=f_{K^{\ast}}\sqrt{15}c_{\overline{10}}$) yields 
about $1.86$ and $1.95$, respectively, whereas the
present value produces $g_{K^{*}n\Theta}=0.96$. We want to mention
that the measurement of the $\Theta^{+}$ photoproduction prefers
smaller values of $g_{K^{*}n\Theta}$~\cite{Nakano:2008ee}. The
detailed analysis of this coupling constant will appear elsewhere.

It is worthwhile to compare closely the present results for the
predicted values of $M_{\Theta^+}$ and $M_{N^*}$ with those of
Refs.~\cite{Diakonov:1997mm,Ellis:2004uz}. For this comparison, let us
turn off the effects of isospin symmetry breaking. Then, the only
relevant parameter is $m_{\mathrm{s}}\delta_3$ for the masses of the
baryon antidecuplet, as shown in Eq.~(\ref{eq:antidecM}). The
corresponding results are obtained, respectively, as  
$m_{\mathrm{s}}\delta_3=177$ MeV (Diakonov et al.),
$m_{\mathrm{s}}\delta_3=108$ MeV (Ellis et al.), and 
$m_{\mathrm{s}}\delta_3=171$ MeV from the present work. 
Using the mass of $\Theta^+$ measured by the LEPS collaboration,  
i.e. $M_{\Theta^+} =1524$ MeV, one finds the $N^*$ masses,
respectively, as follows: $M_{N^*}=1700$ MeV (Diakonov et al.),
$M_{N^*}=1631$ MeV (Ellis et al.), $M_{N^*}=1694$ MeV for this
work. The experimental data for the $N^*$ mass
$M_{N^*}=(1685\pm12)$ MeV~\cite{Kuznetsov:2008ii} being considered,
the present result turns out to be quite comparable to it. We will
show that the results will be improved later with the effects of
isospin symmetry breaking switched on.  

\begin{table}[th]
 \centering \caption{Reproduced masses of the baryon octet. The
   experimental data of octet baryons are taken from the Particle 
Data Group (PDG).} 
\begin{tabular}{cccccc}
\hline 
\multicolumn{2}{c}{Mass {[}MeV{]}} & $T_{3}$  & $Y$  
& Exp. {[}Inputs{]}  & Numerical results
\tabularnewline
\hline 
$M_{N}$  & $\begin{array}{c}
p\\
n
\end{array}$  & $\begin{array}{c}
\;\;1/2\\
-1/2
\end{array}$  & $\;\;1$  & $\begin{array}{c}
938.27203\pm0.00008\\
939.56536\pm0.00008
\end{array}$  & $\begin{array}{c}
938.76\pm3.65\\
940.27\pm3.64
\end{array}$\tabularnewline
\hline 
$M_{\Lambda}$  & $\Lambda$  & $\;\;0$  
& $\;\;0$  & $1115.683\pm0.006$  & $1109.61\pm0.70$
\tabularnewline
\hline 
$M_{\Sigma}$  & $\begin{array}{c}
\Sigma^{+}\\
\Sigma^{0}\\
\Sigma^{-}
\end{array}$  & $\begin{array}{c}
\;\;1\\
\;\;0\\
-1
\end{array}$  & $\;\;0$  & $\begin{array}{c}
1189.37\;\pm0.07\\
1192.642\pm0.024\\
1197.449\pm0.030
\end{array}$  & $\begin{array}{c}
1188.75\pm0.70\\
1190.20\pm0.77\\
1195.48\pm0.71
\end{array}$\tabularnewline
\hline 
$M_{\Xi}$  & $\begin{array}{c}
\Xi^{0}\\
\Xi^{-}
\end{array}$  & $\begin{array}{c}
\;\;1/2\\
-1/2
\end{array}$  & $-1$  & $\begin{array}{c}
1314.83\pm0.20\\
1321.31\pm0.13
\end{array}$  & $\begin{array}{c}
1319.30\pm3.43\\
1324.52\pm3.44
\end{array}$\tabularnewline
\hline 
\end{tabular}\label{tab:Octm1} 
\end{table}
In Table~\ref{tab:Octm1}, the reproduced masses of the baryon octet
are listed. In Table~\ref{tab:decm1}, the predicted results of the
decuplet masses are listed. Those of the $\Sigma^{\ast}$ and $\Xi^{\ast}$ are in
remarkable agreement with the data within $0.5\,\%$. In
Table~\ref{tab:antim1} we present the results for the masses of the
baryon antidecuplet. 

\begin{table}[th]
 \centering \caption{Predicted masses of the baryon decuplet. The
   experimental data of decuplet baryons are taken from the Particle Data
   Group (PDG).} 
\begin{tabular}{cccccc}
\hline 
\multicolumn{2}{c}{Mass {[}MeV{]}} 
& $T_{3}$  
& $Y$  
& Experiment \cite{PDG}  
& Predictions
\tabularnewline
\hline 
$M_{\Delta}$  
& $\begin{array}{c}
\Delta^{++}\\
\Delta^{+}\\
\Delta^{0}\\
\Delta^{-}
\end{array}$  
& $\begin{array}{c}
\;\;3/2\\
\;\;1/2\\
-1/2\\
-3/2
\end{array}$  
& $\;\;1$  
& $1231-1233$  
& $\begin{array}{c}
1248.54\pm3.39\\
1249.36\pm3.37\\
1251.53\pm3.38\\
1255.08\pm3.37
\end{array}$\tabularnewline
\hline 
$M_{\Sigma^{\ast}}$  
& $\begin{array}{c}
\Sigma^{\ast+}\\
\Sigma^{\ast0}\\
\Sigma^{\ast-}
\end{array}$  
& $\begin{array}{c}
\;\;1\\
\;\;0\\
-1
\end{array}$  
& $\;\;0$  
& $\begin{array}{c}
1382.8\pm0.4\;\\
1383.7\pm1.0\\
1387.2\pm0.5\,
\end{array}$  
& $\begin{array}{c}
1388.48\pm0.34\\
1390.66\pm0.37\\
1394.20\pm0.34
\end{array}$
\tabularnewline
\hline 
$M_{\Xi^{\ast0}}$  
& $\begin{array}{c}
\Xi^{\ast0}\\
\Xi^{\ast-}
\end{array}$  
& $\begin{array}{c}
\;\;1/2\\
-1/2
\end{array}$  
& $-1$  
& $\begin{array}{c}
1531.80\pm0.32\\
1535.0\;\pm0.6\;\,
\end{array}$  
& $\begin{array}{c}
1529.78\pm3.38\\
1533.33\pm3.37
\end{array}$
\tabularnewline
\hline 
$M_{\Omega^{-}}^{\star}$  
& $\Omega^{-}$  
& $0$  & $-2$  
& $1672.45\pm0.29$  
& Input
\tabularnewline
\hline 
\end{tabular}\label{tab:decm1} 
\end{table}

\begin{table}[h]
 \centering \caption{Predicted masses of the baryon antidecuplet.}
\begin{tabular}{cccccc}
\hline 
\multicolumn{2}{c}{Mass } 
& $T_{3}$  
& $Y$  
& Experiment  
& Predictions
\tabularnewline
\hline 
$M_{\Theta^{+}}$  
& $\Theta^{+}$  
& $\;\;0$  
& $\;\;2$  
& $1524\pm 5$\cite{Nakano:2008ee}  
& Input
\tabularnewline
\hline 
$M_{N^{\ast}}$  
& $\begin{array}{c}
p^{\ast}\\
n^{\ast}
\end{array}$  
& $\begin{array}{c}
\;\;1/2\\
-1/2
\end{array}$  
& $\;\;1$  
& ${\displaystyle 1686\pm 12}$
\cite{Kuznetsov:2008ii}
& $\begin{array}{c}
1688.18\pm10.53\\
1692.16\pm10.53
\end{array}$\tabularnewline
\hline 
$M_{\Sigma_{\overline{10}}}$  
& $\begin{array}{c}
\Sigma_{\overline{10}}^{+}\\
\Sigma_{\overline{10}}^{0}\\
\Sigma_{\overline{10}}^{-}
\end{array}$  
& $\begin{array}{c}
\;\;1\\
\;\;0\\
-1
\end{array}$  
& $\;\;0$  
&  
& $\begin{array}{c}
1852.35\pm10.00\\
1856.33\pm10.00\\
1858.95\pm10.00
\end{array}$\tabularnewline
\hline 
$M_{\Xi_{3/2}}$  
& $\begin{array}{c}
\Xi_{3/2}^{+}\\
\Xi_{3/2}^{0}\\
\Xi_{3/2}^{-}\\
\Xi_{3/2}^{--}
\end{array}$  
& $\begin{array}{c}
\;\;3/2\\
\;\;1/2\\
-1/2\\
-3/2
\end{array}$  
& $\;\;-1$  
&  
& $\begin{array}{c}
2016.53\pm10.53\\
2020.51\pm10.53\\
2023.12\pm10.53\\
2024.37\pm10.53
\end{array}$\tabularnewline
\hline 
\end{tabular}
\label{tab:antim1} 
\end{table}

In this work, we use the mass of the $\Theta^{+}$ taken from
Ref.~\cite{Nakano:2008ee} as input. Though the NA49 data of
$\Xi_{3/2}^{--}(1862)$ ~\cite{Alt:2003vb} is still under debate, we
can compare the present results with that. As shown in
Table~\ref{tab:antim1}, the results seem to be quite larger than the
NA49 data. Note that, however, A recent analysis of
Ref.~\cite{Goeke:2009ae} yields the mass ranges of
$\Sigma_{\overline{10}}$ and $\Xi_{3/2}$ as  
\begin{equation}
M_{\Sigma_{\overline{10}}}
\;=\;1795\,\mathrm{MeV}-1830\,\mathrm{MeV},
\;\;\;\; M_{\Xi_{3/2}}\;=\;1900\,\mathrm{MeV}-1970\,\mathrm{MeV},
\label{eq:S10barX10bar}
\end{equation}
 which are comparable to the present results.

\section{Summary and conclusion}

In the present work, we have investigated the mass splittings of the
SU(3) baryons within the framework of an SU(3) chiral soliton model,
taking into account SU(3) and isospin symmetry breakings due to the
electromagnetic self-interactions as well as hadronic isospin mass
differences. We found various mass relations of the baryon octet,
decuplet, and antidecuplet. In particular, we obtained the generalized
Gell-Mann-Okubo mass formulae that are well satisfied with the
experimental data. We also derived the Coleman-Glashow mass formula
and Guadagnini mass relation. In addition, similar mass relations in
the baryon antidecuplet were presented. 

In order to determine the unknown model parameters $\alpha$, $\beta$,
and $\gamma$, we employed the existing experimental data for the
baryon octet, the $\Omega^{-}$, and the $\Theta^{+}$. We then
performed the minimization of the $\chi^{2}$. The second moment of
inertia $I_{2}$ was also found, which is an essential key to explain
the mass splittings within the baryon antidecuplet. Moreover, the
pion-nucleon sigma term was determined to be $\Sigma_{\pi
  N}=(36.4\pm3.9)$ MeV. The present results of the $\Sigma^{*}$ and
$\Xi^{*}$ masses were in remarkable agreement with the experimental
data. It indicates that the mass of the $\Theta^{+}$ used as input in
the present scheme is rather compatible with existing experimental
data for the baryon octet and decuplet. 

The present work is distinguished from the previous
studies~\cite{Diakonov:1997mm,Ellis:2004uz} based on the chiral
soliton model, which also deal with the mass splittings of the SU(3)
baryons. The second moment of inertia $I_{2}$ plays a crucial role in
explaining the heavier masses of the baryon antidecuplet, compared to
those of the octet and decuplet. However, it was not possible to fix
it unambiguously in previous works, so that results of the model
calculations had to be used. Moreover, since the $\Sigma_{\pi N}$ was
not uniquely known empirically, some ambiguities were inevitable in
previous analyses. While Refs.~\cite{Diakonov:1997mm,Ellis:2004uz}
used the experimental data for the baryon octet, they did not consider
isospin symmetry breaking, so that they were unable to incorporate
whole experimental information. 

In the present work, we were able to fix all model parameters by using 
the experimental data for the masses of the baryon octet and parts
of the baryon decuplet and antidecuplet, because effects of isospin
symmetry breaking have been fully taken into account. Thus, we have
produced the masses of the baryon antidecuplet as well as of the
decuplet without any further adjustable parameter.

While we determined the masses of the baryon decuplet and
antidecuplet, we have not considered in the present work the
corresponding decay widths which are very important to understand
those baryons. In the previous works in the chiral soliton models, one
of the parameters for the decay width of the $\Theta^{+}$ has been
taken from the model calculations. Moreover, the effects of SU(3)
symmetry breaking have never been fully considered. In order to
calculate the widths of the antidecuplet systematically, we have to
fix all relevant parameters, using the experimental data for
axial-vector constants as well as hyperon semileptonic decays. The
corresponding investigation is under way. 
\section*{Acknowledgments}
The authors are grateful to the late Klaus Goeke for his support and
contribution to the early stage of the present work. They are also
thankful to M.V. Polyakov for his interest in this work and
constructive suggestions. They express their gratitude to J. Franklin
for constructive criticism of various mass relations. Gh.-S. Yang
expresses his gratitude to M.~Prasza\l{}owicz and S.~i. Nam for
valuable discussions and comments. The present work is supported by
Inha University Research Grant.


\begin{thebibliography}{99}
\bibitem{Jezabek:1987ns} M.~Prasza{\l{}}owicz, in Proceedings
of the Workshop on Skyrmions and Anomalies, Krakow, Poland, 1987,
Eds. M.~Jezabek and M.~Prasza{\l{}}owicz, (World Scientific, Singapore,
1987).

\bibitem{Diakonov:1997mm} D.~Diakonov, V.~Petrov, and M.~V.~Polyakov,
Z. Phys. A \textbf{359} (1997), 305.

\bibitem{Praszalowicz:2003ik} M.~Praszalowicz, 
 Phys. Lett. B \textbf{575} (2003), 234. 


\bibitem{Nakano:2003qx} T.~Nakano \textit{et al.} {[}LEPS Collaboration{]},
Phys. Rev. Lett. \textbf{91}(2003), 012002 . 


\bibitem{Battaglieri:2005er} M.~Battaglieri \textit{et al.} {[}CLAS
Collaboration{]}, 
 Phys.\ Rev.\ Lett.\ \textbf{96} (2006), 042001. 


\bibitem{McKinnon:2006zv} B.~McKinnon \textit{et al.} {[}CLAS Collaboration{]},
 Phys. Rev. Lett. \textbf{96} (2006), 212001. 


\bibitem{Niccolai:2006td} S.~Niccolai \textit{et al.} {[}CLAS Collaboration{]},
 Phys. Rev. Lett. \textbf{97} (2006), 032001. 


\bibitem{DeVita:2006ha} R.~De Vita \textit{et al.} {[}CLAS Collaboration{]},
 Phys. Rev. D \textbf{74} (2006), 032001. 


\bibitem{Barmin:2006we} V.~V.~Barmin \textit{et al.} {[}DIANA Collaboration{]},
 Phys. Atom. Nucl. \textbf{70} (2007), 35. 


\bibitem{Barmin:2009cz} V.~V.~Barmin \textit{et al.} {[}DIANA Collaboration{]},
 Phys.\ Atom.\ Nucl.\ \textbf{73} (2010), 1168. 


\bibitem{new_SVD} A. Aleev and {[}SVD Collaboration{]}, hep-ex/0509033.

\bibitem{Hotta:2005rh} T.~Hotta {[}LEPS Collaboration{]}, 
 Acta Phys. Polon. B \textbf{36} (2005), 2173. 


\bibitem{Miwa:2006if} K.~Miwa \textit{et al.} {[}KEK-PS E522 Collaboration{]},
 Phys.\ Lett.\ B \textbf{635} (2006), 72. 


\bibitem{SVD2:2008} A. Aleev \textit{et al.} {[}SVD Collaboration{]},
 arXiv:0803.3313 {[}hep-ex{]}

\bibitem{Nakano:2008ee} T.~Nakano \textit{et al.} {[}LEPS Collaboration{]},
 Phys.\ Rev.\ C \textbf{79} (2009), 025210. 


\bibitem{Kuznetsov:2004gy} V.~Kuznetsov {[}GRAAL Collaboration{]},
arXiv:hep-ex/0409032. 


\bibitem{Kuznetsov:2006de} V.~Kuznetsov \textit{et al.}, 
 arXiv:hep-ex/0601002. 


\bibitem{Kuznetsov:2006kt} V.~Kuznetsov {[}GRAAL Collaboration{]},
 Phys.\ Lett.\ B \textbf{647} (2007), 23. 


\bibitem{Fix:2007st} A.~Fix, L.~Tiator and M.~V.~Polyakov, 
 Eur.\ Phys.\ J.\ A \textbf{32} (2007), 311. 


\bibitem{Diakonov:2003jj} D.~Diakonov and V.~Petrov, 
 Phys.\ Rev.\ D \textbf{69} (2004), 094011. 


\bibitem{Arndt:2003ga} R.~A.~Arndt, Y.~I.~Azimov, M.~V.~Polyakov,
I.~I.~Strakovsky and R.~L.~Workman, 
 Phys.\ Rev.\ C \textbf{69} (2004), 035208. 


\bibitem{Polyakov:2003dx} M.~V.~Polyakov and A.~Rathke, 
 Eur.\ Phys.\ J.\ A \textbf{18} (2003), 691. 


\bibitem{Kim:2005gz} H.-Ch.~Kim, M.~Polyakov, M.~Prasza\l{}owicz,
G.~S.~Yang and K.~Goeke, 
Phys.\ Rev.\ D \textbf{71} (2005), 094023.

\bibitem{Kuznetsov:2007dy} V.~Kuznetsov, M.~Polyakov, T.~Boiko,
J.~Jang, A.~Kim, W.~Kim and A.~Ni, 
 arXiv:hep-ex/0703003. 


\bibitem{Kuznetsov:2008gm} V.~Kuznetsov \textit{et al.}, 
 arXiv:0801.0778 {[}hep-ex{]}. 


\bibitem{Kuznetsov:2010as} V. Kuznetsov \textit{et al.}, 
arXiv:1003.4585 {[}hep-ex{]}.

\bibitem{Kuznetsov:2008hj} V. Kuznetsov \textit{et al.}, 
Acta Phys. Polon. B \textbf{39} (2008), 1949.

\bibitem{Kuznetsov:2008ii} V. Kuznetsov and M. V. Polyakov, 
JETP Lett. \textbf{88} (2008), 347 {[}arXiv:0807.3217 {[}hep-ph{]}{]}.

\bibitem{Bartalini:2007fg} O.~Bartalini \textit{et al.} {[}The GRAAL
collaboration{]}, 
 Eur.\ Phys.\ J.\ A \textbf{33} (2007), 169. 


\bibitem{CBELSA} D. Elsner and {[}CBELSA Collaboration{]}, Eur. Phys.
J. A\textbf{33} (2007), 147.

\bibitem{Azimov:2005jj} Y.~Azimov, V.~Kuznetsov, M.~V.~Polyakov
and I.~Strakovsky, 
 Eur.\ Phys.\ J.\ A \textbf{25} (2005), 325.

\bibitem{Choi:2005ki} K.~S.~Choi, S.~i.~Nam, A.~Hosaka and H.-Ch.~Kim,
Phys. \ Lett.\ B \textbf{636} (2006), 253. 


\bibitem{Choi:2007gy} K.~S.~Choi, S.~i.~Nam, A.~Hosaka and H.-Ch.~Kim,
Jour. Phys. G \textbf{36} (2009), 015008.

\bibitem{non_strange_partner2} I.I. Strakovsky, R.A. Arndt, Ya.I.
Azimov, M.V. Polyakov and R.L. Workman, AIP Conf.~Proc., \textbf{775} (2005),
41.

\bibitem{Christov:1993ny} C.~V.~Christov, A.~Blotz, K.~Goeke,
P.~Pobylitsa, V.~Petrov, M.~Wakamatsu and T.~Watabe, 
Phys. Lett. B \textbf{325} (1994), 467. 


\bibitem{Blotz:1994wi} A.~Blotz, M.~Prasza{\l{}}owicz and K.~Goeke,
 Phys. Rev. D \textbf{53} (1996), 485.

\bibitem{Blotz:1992br} A. Blotz, K.Goeke, N. W. Park, D. Diakonov,
V. Petrov and P. V. Pobylitsa,~
Phys. Lett. B \textbf{287} (1992), 29.

\bibitem{Blotz:1992pw} A.~Blotz, D.~Diakonov, K.~Goeke, N.~W.~Park,
V.~Petrov and P.~V.~Pobylitsa,~
 Nucl. Phys. A \textbf{555} (1993), 765.

\bibitem{Walliser:1992am} H.~Walliser, in Proceedings of the Workshop
on Baryons as Skyrme Solitons, Siegen, Germany, 1992, Ed. G. Holzwarth
(World Scientific, Singapore, 1992).

\bibitem{Walliser:1992vx} H.~Walliser, 
 Nucl. Phys. A \textbf{548} (1992), 649. 


\bibitem{PDG} K. Nakamura \textit{et al}. {[}Particle Data Group{]},
J. Phys. G \textbf{37} (2010), 075021.

\bibitem{Yang:2010id} G.~S.~Yang, H.-Ch.~Kim and M.~V.~Polyakov,
{[}arXiv:1009.5250 {[}hep-ph{]}{]}, Phys.~Lett.~B. (2010) in press.

\bibitem{Ellis:2004uz} J.~R.~Ellis, M.~Karliner and M.~Praszalowicz,
 JHEP \textbf{0405} (2004), 002. 


\bibitem{Praszalowicz:1992gn} M.~Prasza{\l{}}owicz, A.~Blotz
and K.~Goeke, 
 Phys.\ Rev.\ D \textbf{47} (1993), 1127. 


\bibitem{Blotz:1994pc} A.~Blotz, K.~Goeke and M.~Prasza\l{}owicz,
Acta Phys. Polon. B \textbf{25} (1994), 1443 .

\bibitem{Christov:1995vm} C.~V.~Christov \textit{et al.},
Prog. Part. Nucl. Phys. \textbf{37} (1996), 91.

\bibitem{Witten:1983tx} E.~Witten, 
 ~Nucl. Phys. B \textbf{223} (1983), 433 .

\bibitem{Guadagnini:1983uv} E. Guadagnini,~ Nucl. Phys . B \textbf{236} (1984),
35.

\bibitem{Jain:1984gp} S. Jain and S. R. Wadia,~ Nucl. Phys. B \textbf{258} (1985),
713.

\bibitem{Pobylitsa:1992bk} P.~V.~Pobylitsa, E.~Ruiz Arriola, T.~Meissner,
F.~Grummer, K.~Goeke and W.~Broniowski, ~J.Phys. G \textbf{18} (1992),
1455.

\bibitem{Kim:1998gt} H.-Ch.~Kim, M.~Prasza\l{}owicz, M.~V.~Polyakov
and K.~Goeke, 
 Phys. Rev. D \textbf{58} (1998), 114027.

\bibitem{Weinberg:1959zzb} S.\ Weinberg and S.\ B.\ Treiman, Phys.
Rev. \textbf{116} (1959), 465. 


\bibitem{Dashen} R. Dashen, Phys.\ Rev.\ \textbf{183} (1969), 1245.

\bibitem{Gasser:1982ap} J.~Gasser and H.~Leutwyler, 
 ~Phys. Rept. \textbf{87} (1982), 77.

\bibitem{Coleman:1961jn} S.~R.~Coleman and S.~L.~Glashow, 
 ~Phys. Rev. Lett. \textbf{6} (1961), 423.

\bibitem{GellMann:1962xb} M.~Gell-Mann, 
~ Phys. Rev. \textbf{125} (1962), 1067.

\bibitem{Okubo:1961jc} S.~Okubo, 
 ~Prog. Theor. Phys. \textbf{27} (1962), 949.

\bibitem{Morpurgo:1991if} G.~Morpurgo, 
Phys.~Rev.~Lett. \textbf{68} (1992), 139.

\bibitem{Ledwig:2008rw} T.~Ledwig, H.-Ch.~Kim and K.~Goeke, 
Phys. Rev. D \textbf{78} (2008), 054005.

\bibitem{Gasser:1990ce} J.~Gasser, H.~Leutwyler and M.~E.~Sainio,
 Phys. Lett. B \textbf{253} (1991), 252. 


\bibitem{Pavan:2001wz} M.~M.~Pavan, I.~I.~Strakovsky, R.~L.~Workman
and R.~A.~Arndt, 
 PiN Newslett. \textbf{16} (2002), 110. 


\bibitem{Schweitzer:2003fg} P.~Schweitzer, 
 Eur. Phys. J. A \textbf{22} (2004), 89. 


\bibitem{Ledwig:1900ri} T.~Ledwig, H.~C.~Kim and K.~Goeke, 
 Nucl.\ Phys.\ A \textbf{811} (2008), 353.


\bibitem{Alt:2003vb} C. Alt \textit{et al}. {[}NA49 Collaboration{]},
Phys. Rev. Lett. \textbf{92} (2004), 042003.

\bibitem{Goeke:2009ae} K. Goeke, M. V. Polyakov and M. Prasza\l{}owicz,
{[}arXiv:0912.0469 {[}hep-ph{]}{]}. 
\end{thebibliography}
\end{document}